# Practical Solutions for 400 Gbit/s Data Center Transmission

Annika Dochhan[1], Nicklas Eiselt[1,2], Jinlong Wei[1], Helmut Griesser[3], Michael Eiselt[1], Juan José Vegas Olmos[2], Idelfonso Tafur Monroy[2], Jörg-Peter Elbers[3]

*1 ADVA Optical Networking SE, Maerzenquelle 1-3, 98617 Meiningen, Germany*
*2 Technical University of Denmark (DTU), Department of Photonics Engineering, Ørsteds Plads, Build. 343, DK-2800*
*3 ADVA Optical Networking SE, Fraunhoferstr. 9a, 82152 Martinsried, Germany*
*ADochhan@advaoptical.com*

**Abstract:** We review three solutions for low-cost data center interconnects with a target reach of up to 80 km. Directly detected DMT, PAM-4 and multi-band CAP are promising modulation schemes, enabling 400 Gbit/s by combining eight channels of 56 Gbit/s.

## 1. Introduction

The massive growth of internet traffic of up to 25 % each year demand new solutions for high speed intra- and inter-data center interconnect (DCI) applications [1]. For intra-data center applications standardization efforts from the IEEE P802.3bs Task Force are still ongoing, but the road seems clear with the selection of 4-level pulse amplitude modulation (PAM-4) for 400 GbE transmission over a single mode fiber (SMF) [2]. For such distances, footprint, power consumption and cost-effectiveness are of primary concern, while fiber exhaust and long reach are less; favoring intensity modulation and direct detection (IM/DD) and the transmission wavelength of 1.3 µm. Inter-data center applications, however, tend to cover geographic areas of up to 80 km, which are beyond the IEEE client optic standards. Such applications demand multi-Tb/s (multiples of 400 Gbit/s) dense wavelength division multiplexed (DWDM) links and the 1.5 µm transmission window to enable the use of erbium doped fiber amplifiers (EDFAs). Here, footprint, power consumption and latency are further critical parameters, which need to be considered. Coherent optics, such as DP-16-QAM or DP-64-QAM (dual polarization quadrature amplitude modulation), offer very high spectral efficiency with up to 200 Gbit/s or even 400 Gbit/s per wavelength, respectively. However, this spectral efficiency is not required for such an application and their cost and power consumption might not satisfy the stringent requirements in the near future. Therefore, there has been an increased interest to leverage the cost efficient client optics IM/DD approach for inter-DCI applications. Extensive investigations have been made with PAM-4 and discrete multi-tone (DMT) for such a scenario resulting already in real-time transmission experiments [3]–[5], but also carrier-less amplitude and phase (CAP) modulation was proposed for this purpose and showed noticeable performance. In this paper, the latest achievements of the three mentioned modulation formats are reviewed. Afterwards, an experimental comparison at a channel net-rate of 56 Gbit/s is shown.

## 2. IM/DD Solutions for 400G Inter-DCI Applications

To achieve 400G based on IM/DD, multi-wavelength solutions are mandatory, e.g. a superchannel comprising eight wavelengths carrying 56 Gbit/s each or four wavelengths with 112 Gbit/s each. Considering current bandwidth limitations of low-cost electrical components, 56 Gbit/s seems to be the short-term solution. However, latest achievements also show the feasibility of 100 Gbit/s per λ and beyond, when high-bandwidth components are used.

Multicarrier formats, such as DMT, have the ability to assign modulation format and power to each subcarrier (bit and power loading: BL, PL) individually, so that component bandwidth limitations and power fading due chromatic dispersion (CD) can be effectively compensated; but may suffer from a high peak-to-average power ratio (PAPR). The extensive investigations to bridge 80 km SSMF at 1.5 µm using 100 Gb/s DMT per λ without the use of an optical dispersion compensation module (DCM), resulted in different flavours of DMT, but always showing the necessity of using either single-sideband (SSB) or vestigial sideband (VSB) transmission [6]–[8]. Advanced signal processing such as nonlinear Volterra equalization [6], subcarrier-interference cancellation [7], [9], trellis coded modulation [6], DFT spreading [8], asymmetrical clipping [10] or pre-equalization [8] showed further performance improvement. In addition, an external modulator such as a Mach-Zehnder modulator (MZM), dual-drive MZM (DDMZM), IQ-MZM or an electro-absorption modulator (EAM) seems to be required for such a scenario [8]-[12]. Latest demonstrations with e.g. a vertical-cavity surface-emitting laser (VCSEL) or a directly modulated laser (DML) revealed limited transmission distance or data rate [13]–[15].

CAP is another strong contender for such an application scenario, as it offers good performance with respect to ease of implementation, CD tolerance and DSP complexity. However, it has been investigated mostly for short reach application and/or O-band transmission [16]–[18], multi-mode systems [19], passive optical network (PON) systems [20] and only lately attracted attention for 80 km SSMF transmission at 1.5 µm [21]–[23]. Among the different fla-vors, multi-band CAP seems to be the most promising one, as it offers, similar to DMT, the ability to do BL and PL for each sub-band individually. Using VSB, transmission over up to 80 km of SSMF at 56 Gbit/s without an optical DCM was shown in [23].

IM/DD PAM-4 is known to be very sensitive to CD, which limits its transmission reach at 1.5 µm at higher data rates even when very complex maximum-likelihood sequence estimation (MLSE) is applied [24]. Therefore, and because of standardization efforts from the IEEE, PAM-4 has been extensively investigated for short reach applications and less for 80 km. However, if optical CD compensation is allowed in the system – the costs of the DCM are shared between all channels and are therefore negligible – PAM-4 is an interesting modulation format also for 80 km due to its simplicity and the good performance [3], [25], [26]. PAM-4 does not necessarily require high-resolution digital-to-analog converters (DACs), which frequently suffer from limited bandwidth, but could be generated using only a 2-bit or 3-bit DAC [27], a simple power-combiner [28] or a segmented modulator [4]. Using these devices, IM/DD PAM-4 at 168 Gbit/s and beyond has been shown already [27]. Similar to CAP or DMT, also PAM-4 was shown in various flavours, such as Nyquist-PAM-4 [29], duobinary PAM-4 [30], quasi-SSB Nyquist PAM-4 [31], PAM-4 with partial response equalization in combination with MLSE [32] and multi-dimensional-coded PAM-4 [33], [34] allowing data rates up to 100 Gbit/s. Again, an external modulator seems to be required for 80 km transmission, while a DML or a VCSEL are limited to short-reach applications.

### 3. Experimental Setup

Fig. 1 shows the system setup. The data was pre-processed offline (Tx DSP) and loaded onto the memory of a DAC running at 84 GS/s for PAM-4 and DMT and at 80 GS/s for multi-band CAP. The differential output of the DAC was directly connected to a DDMZM with ~27 GHz bandwidth. A 50-GHz-channel grid WDM multiplexer (MUX) with a passband width of 39 GHz was applied to account for the bandwidth limitations that would occur in a WDM system. The MUX was followed by a booster EDFA whose output power (and thus the fiber launch power) was controlled by a variable optical attenuator (VOA). 80 km of SSMF was followed by another VOA to vary the receiver optical signal-to-noise ratio (OSNR) and a pre-amplifier EDFA. For PAM-4, a DCM was inserted between the two stages of this EDFA. At the EDFA output, the OSNR was determined with the help of an optical spectrum analyzer (OSA). A de-multiplexer (DEMUX) with 50-GHz-channel grid again emulated the bandwidth limitation, but also reduced the noise to the receiver, which is a combination of PIN diode and trans-impedance amplifier (PIN/TIA). The input power to the PIN/TIA was controlled by another VOA. After amplification, the signal was captured by a digital-to-analog converter (ADC), which ran at the same rate as the DAC, and sent to the computer for offline demodulation and decoding.

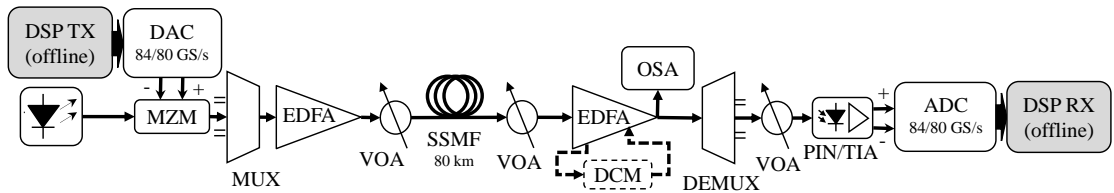

**Fig. 1:** System setup for all considered modulation formats. For PAM-4 and DMT the DAC and ADC run at 84 GS/s, wheras CAP used 80 GS/s. The dispersion compensating module (DCM) was only needed for PAM-4.

For DMT and CAP, the option of VSB filtering to improve chromatic dispersion tolerance was also investigated. In the DSB case, after direct detection, the upper and the lower part of the optical spectrum mix and lead to spectral fading at defined frequencies (power fading). Since all information is present in both bands, SSB transmission is one option to avoid this effect. True SSB transmission requires additional effort at the Tx side (at least a very steep filter) and leads to additional non-linear mixing products after detection. Therefore, we chose to filter out one of the sidebands partially with the MUX/DEMUX and suitable detuning of the transmit laser [35], [36]. The optimum shift for both formats was found to be 20 GHz with a tolerance of 5-6 GHz.

The flow charts of the DSP for each modulation format are shown in Fig. 2. A detailed description of each block can be found in [23], [35], [36]. For the comparison all formats run at 56 Gbit/s. PAM-4 used a $T/2$-spaced feedforward equalizer (FFE) with 11 filter taps at the Rx side to compensate for bandwidth impairments. The DMT signal was created with a 512-point FFT and uses a cyclic prefix (CP) of eight samples. The multi-band CAP signal consisted of 12 sub-bands, each with a symbol rate of 2 GBd. For CAP, to enable completely blind processing at the Rx side, differential QAM constellation encoding/decoding was applied. A multi-modulus algorithm-based, $T/2$-spaced 14-tap FFE was used for each band to recover the signal. DMT and multi-band CAP used bit and power loading according to a previously estimated channel SNR to account for bandwidth limitations and other impairments.

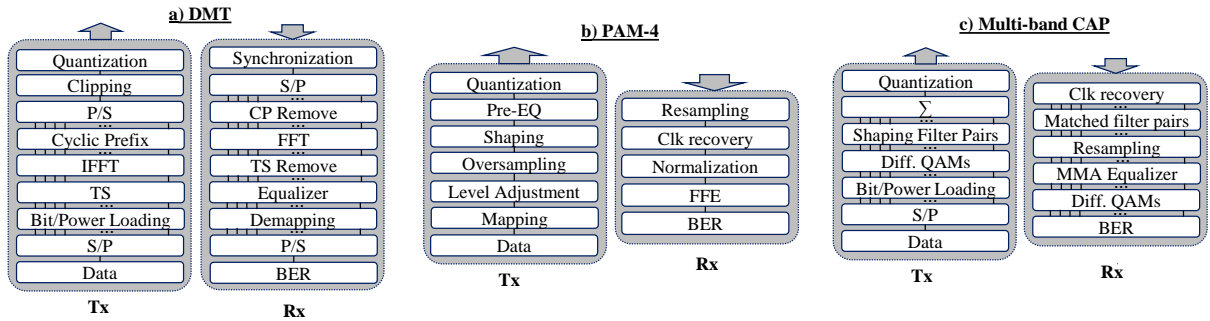

**Fig. 2:** DSP flow charts of DMT (a), PAM4 (b) and multi-band CAP (c).

## 4. Results

Fig. 3 shows the results for back-to-back (b2b) and transmission over 80 km. A table summarizes the required OSNR (ROSNR) for a BER of $3.8 \cdot 10^{-3}$, which can ensure a residual error rate below $10^{-15}$ if an appropriate forward error correction (FEC) is applied. The fiber launch and the receiver input powers were optimized for all formats [23], [35], [36]. For PAM-4 a CD compensating fiber is employed at the receiver side, whereas DMT and CAP are more dispersion tolerant, especially in the case of VSB transmission. PAM-4 shows the best results with 23.5 dB ROSNR in b2b case and only 1 dB penalty for 80 km transmission. DMT and CAP require roughly 3-4 dB more OSNR in both cases if VSB is considered. To a limited extent, DSB DMT is still capable of accounting for the dispersion induced power fading, leading to a required OSNR of 32.2 dB for 80 km transmission. Since multi-band CAP with 14 bands cannot follow the dispersion notches as precise as DMT, DSB transmission over 80 km does not yield sufficiently low BER values. Since the systems for CAP and DMT do not require CD compensation, the maximum available system OSNR could be higher than that for PAM-4.

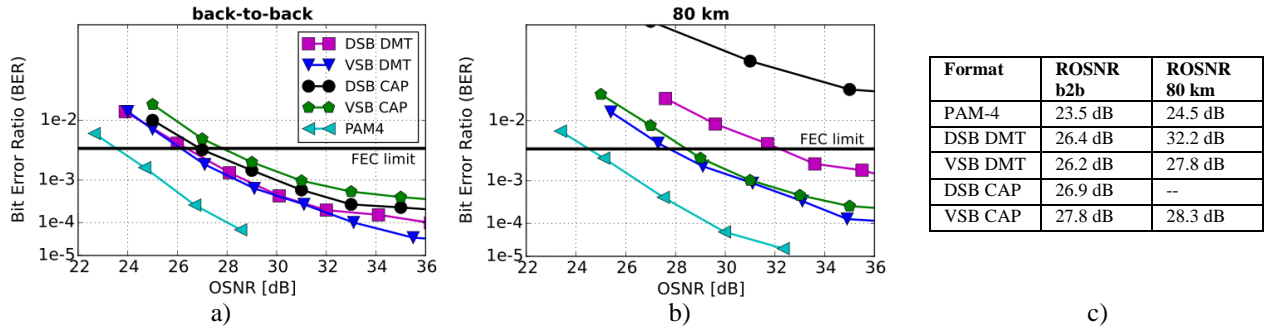

| Format | ROSNR b2b | ROSNR 80 km |
|---|---|---|
| PAM-4 | 23.5 dB | 24.5 dB |
| DSB DMT | 26.4 dB | 32.2 dB |
| VSB DMT | 26.2 dB | 27.8 dB |
| DSB CAP | 26.9 dB | -- |
| VSB CAP | 27.8 dB | 28.3 dB |

**Fig. 3:** BER vs. OSNR for all formats b2b and after transmission over 80 km. In contrast to CAP and DMT, PAM-4 transmission over 80 km requires optical dispersion compensation. Table: Required OSNR for a BER = $3.8 \cdot 10^{-3}$.

## 5. Acknowledgement

The work has been partially funded by the European Union Marie Curie project ABACUS and CEEOLAN and by the German ministry of education and research (BMBF) in project SpeeD under contract 13N1374.